\documentclass[aps]{revtex4-1}

\usepackage{graphicx}

\begin{document}

\title{Thermodynamic analysis of the Quantum Critical behavior of Ce-lattice compounds}

\author{Julian G. Sereni}
\address{Div. Bajas Temperaturas, CAB-CNEA, CONICET, 8400 Bariloche, Argentina}

\date{\today}

\begin{abstract}

{A systematic analysis of low temperature magnetic phase diagrams
of Ce compounds is performed in order to recognize the
thermodynamic conditions to be fulfilled by those systems to reach
a quantum critical regime and, alternatively, to identify other
kinds of low temperature behaviors. Based on specific heat ($C_m$)
and entropy ($S_m$) results, three different types of phase
diagrams are recognized: i) with the entropy involved into the
ordered phase ($S_{MO}$) decreasing proportionally to the ordering
temperature ($T_{MO}$), ii) those showing a transference of
degrees of freedom from the ordered phase to a non-magnetic
component, with their $C_m(T_{MO})$ jump ($\Delta C_m$) vanishing
at finite temperature, and iii) those ending in a critical point
at finite temperature because their $\Delta C_m$ do not decrease
with $T_{MO}$ producing an entropy accumulation at low
temperature.

Only those systems belonging to the first case, i.e. with
$S_{MO}\to 0$ as $T_{MO}\to 0$, can be regarded as candidates for
quantum critical behavior. Their magnetic phase boundaries deviate
from the classical negative curvature below $T\approx 2.5$\,K,
denouncing frequent misleading extrapolations down to $T=0$.
Different characteristic concentrations are recognized and
analyzed for Ce-ligand alloyed systems. Particularly, a
pre-critical region is identified, where the nature of the
magnetic transition undergoes significant modifications, with its
$\partial C_m/\partial T$ discontinuity strongly affected by
magnetic field and showing an increasing remnant entropy at $T\to
0$. Physical constraints arising from the third law at $T\to 0$
are discussed and recognized from experimental results.}
\end{abstract}


\maketitle

\section{Introduction}

Magnetic phase diagrams are drown resuming the relevant
characteristics of magnetic systems once their basic thermodynamic
properties are recognized. Their comparison allows to distinguish
between general physical phenomena and the particular behavior of
a single compound. The respective magnetic phase boundaries can be
traced by applying standard external control parameters like
chemical composition ($x$), pressure ($p$) or magnetic field
($B$), able to drive the system into regions of relevant interest.
Among them, those related with magnetic instabilities,
frustrations, critical points, exotic phases became increasingly
attractive because they involve novel phenomena, stressing the
understanding of basic physical concepts.

At low temperature, where thermal and quantum fluctuations compete
in energy, the neighborhood of quantum critical points (QCP)
allows to confront the well established thermodynamical laws with
new experimental evidences of quantum effects. In fact, a QCP is
currently defined as the $T=0$ limit for a second order transition
driven by one of the mentioned non-thermal control parameters
\cite{TVojta}. Despite of its unattainable nature at $T=0$, a QCP
carries a sort of $halo$ produced by quantum fluctuations whose
physical effects are observed at finite temperature. The
phenomenology arising from those low lying energy excitations is
usually identified as 'non-Fermi-liquid' (NFL) behavior
\cite{Stewart}, in contrast to the canonical Fermi-liquid (FL)
observed in non-magnetic systems. One of the most relevant
features of NFL systems is the increasing density of low energy
excitations, manifested as a divergence of thermodynamic
parameters like specific heat and thermal expansion divided
temperature, and magnetic susceptibility when $T\to 0$.
Accordingly, also electrical resistivity deviates from the $T^2$
dependence of a FL showing a typical linear thermal dependence
\cite{Stewart}. Within this region, the classical magnetic phase
transitions (dominated by thermal fluctuations) transform into
quantum phase transitions (QPT) gradually dominated by quantum
fluctuations \cite{HvL}.

In this work, a comparative analysis of the thermodynamic behavior
of Ce-lattice exemplary compounds is carried in order to determine
the conditions upon which quantum critical or alternative
behaviors can be expected. After to go over the magnetic phase
transitions induced by pressure in metallic Ce, an early
encompassing phase diagram performed on Ce-binary compounds is
reviewed. This phase diagram allowed to recognize some relevant
concentrations for Ce-ligand alloyed systems. The second section
is devoted to identify different types of entropy evolutions as
the ordering temperature decreases because in real systems not all
phase boundaries can be extrapolated to zero without basic
thermodynamical principles violation. The distinct properties of
systems fulfilling the conditions to access to a quantum critical
region are presented and discussed in section III, and the
thermodynamic implications of the third law approaching the $T\to
0$ limit are analyzed in section IV. Hereafter, the parameter
$T_{MO}$ refers to the order temperature independently of its
antiferro (AF) or ferromagnetic (FM) character. However, if any
discussed property is only present in AF systems, the phase
transition will be identified by the usual Neel temperature $T_N$.

\subsection{Ce metal phase diagram}

Despite of the significant progress done during the last decades
in the study of low temperature phase diagrams of intermetallic
compounds containing Ce, Yb and U lattices \cite{Stewart}, some
basic phenomena discovered time ago are still under discussion. To
our knowledge, the first evidence of anomalous behavior of the
$4f$-electrons was observed nearly eighty years ago when the
magnetic behavior of CeN was investigated in 1936 \cite{Iandelli}.
Magnetic measurements showed a weak temperature dependence with a
fractional value of the Ce-$4f$ magnetic moment respect to the
expected from the Hund's rule $J=5/2$ angular moment. This
observation suggested the concept of "intermediate valence" (IV)
for the first time in '$f$' elements. Later, on the early sixties,
the phase diagram of Ce metal became the subject of a systematic
study \cite{kosken}. Driven by pressure, it displays a diversity
of phases where structural and magnetic changes are strongly
related. The $\gamma \leftrightarrow \alpha$ structural and
magnetic phase transition became the fingerprint of the
local-itinerant dilemma of Ce-$4f^1$ electrons \cite{Mackintosh}
not yet completely elucidated.

That first order transition is related to the collapse of the Ce
atomic volume (about $15\%$) and shows an end critical point (CP)
at $\approx 600$\,K under a pressure of $\approx 2$\,GPa. Within
the $\alpha$ structure, a superconductive phase develops up to
$T_{sc} \approx 50$\,mK, which jumps up to $T_{sc} = 1.9$\,K in a
second structural collapse between the $\alpha \leftrightarrow
\alpha'$ phases at 4GPa \cite{wittig}. This superconductive $4f$
mediated phase compares in temperature and IV character with the
recently highlighted second superconductive dome of CeCu$_2$Si$_2$
tuned by pressure and claimed to be related to a second QCP in
that compound \cite{Jaccard}. Since it occurs at the edge of the
heavy fermion HF-IV crossover, where the N=2 degeneracy of the
HF-ground state (GS) transforms into a N=6 GS, this critical
region can be regarded as a reminiscence of the $\alpha
\leftrightarrow \alpha'$ transition. Including the CeTIn$_5$
compounds \cite{CeTIn_5} and Zr$Ce$ alloys \cite{ZrCe}, $\approx
2$\,K seems to be an upper limit for $T_{sc}$ already detected
four decades ago.

\begin{figure}
\begin{center}
\includegraphics[angle=0, width=0.5 \textwidth] {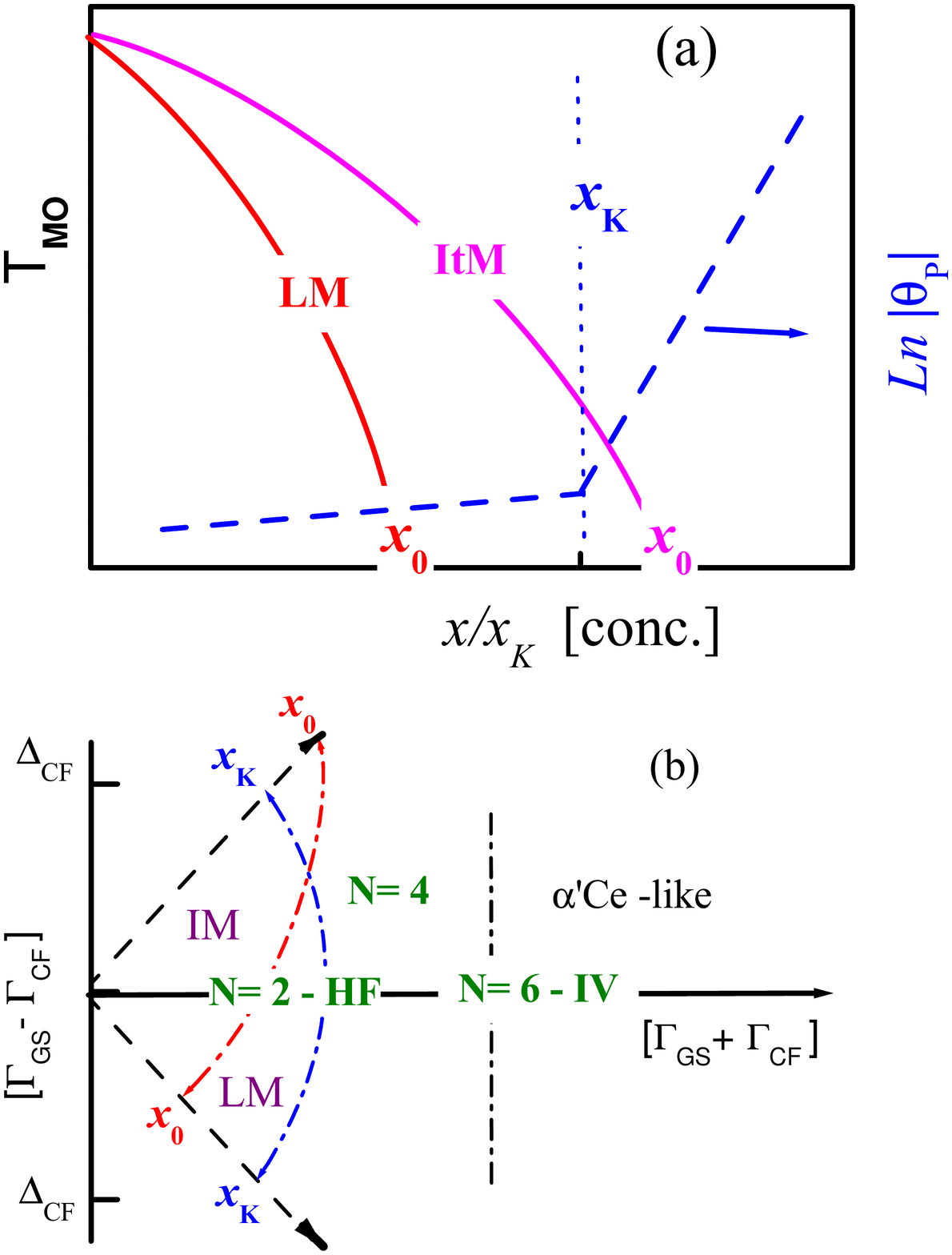}
\end{center}
\caption{Schematic magnetic phase diagrams after
\cite{Systematic}. (a) $T_{MO}$ ordering temperature, $x_0$
concentration of the $T_{MO}(x) \to 0$ extrapolation, $x_K$ onset
of $T_K(x)\propto \theta_P $ increase, LM phase boundary of Ce
compounds with local magnetism ($x_0<x_K$) and ItM with itinerant
magnetic character ($x\geq x_K$). (b) Encompassing phase diagram,
$\Delta_{CF}$ crystal field splitting, $\Gamma_{GS,CF}$ respective
ground and excited levels hybridization strengths, N different GS
degeneracies.} \label{G1}
\end{figure}

\subsection{Low temperature phase diagrams of Ce-lattice compounds}

A turning-point in the description of Ce, Yb and U magnetic phase
diagrams occurred when the competition between {\it on-site} Kondo
screening and {\it inter-site} RKKY interactions was taken into
account by theory \cite{Doniach}. While the former effect weakens
the local $Ce-4f$ moments, the latter provides the long range
interaction favoring the magnetic order. The characteristic
energies of these mechanisms ($k_BT_K$ and $k_BT_R$) can be
compared through their respective dependencies on the
local-conduction band coupling parameter ($g=\delta_F J_{ex}$):
$T_K \propto exp(1/g)$ and $T_R \propto g^2 $, where $\delta_F$ is
the density of states and $J_{ex}(<0)$ the usual exchange
integral. The nature of the GS is therefore established by the
value of $g$, being magnetic ($T_R
>T_K$) for small $g$ and non-magnetic ($T_K >T_R$) for large $g$
values. At the intermediate region, where $T_K$ and $T_R$ are
comparable, both mechanisms compete, weakening the local $4f$
effective moment with the consequent reduction of the ordering
temperature $T_{MO}(g)$.

Although this model contains the basic mechanisms which govern the
magnetic phase boundaries variation, it is evident that a
'one-parameter' description cannot cover the diversity of
behaviors observed applying different control parameters. As
example one can mention the different behavior induced by the
three usual control parameters: $x$, $p$ and $H$, on the specific
heat of a single sample like CeCu$_2$(Si$_{0.9}$Ge$_{0.1}$)$_2$
presente in Fig.15 of Ref.\cite{JPSJ2001}. Furthermore. one may
observe different 'trajectories' of the magnetic phase boundaries
between the ferromagnetic CePd ($T_C=6.5$\,K) and two IV isotypic
compounds CeRh and CeNi, in both cases driven by alloying the
Ce-ligand atom. In the former case: Ce(Pd$_{1-x}$Rh$_x)$, there is
a modification of the {\it chemical potential} whereas in the
later: Ce(Pd$_{1-x}$Ni$_x$), a {\it structural pressure} effect
\cite{Systematic}. Notice the different effect produced by both
control parameters which are usually confused as being equivalent
\cite{FermiEnergy}.

\subsection{Characteristic concentrations in Ce-ligand alloyed systems}

Another limitation for the description of the magnetic phase
diagrams using a single parameter dependence arose from the
comparative analysis performed on seventeen alloyed Ce-binaries
compounds \cite{Systematic}. For such analysis, two characteristic
concentrations were defined: $x_0$, where the magnetic phase
boundary $T_{MO}(x)$ extrapolates to $T=0$, and $x_K$ where the
paramagnetic temperature $\theta_P$ starts to rise significantly.
It is known that $\theta_P \propto T_K$ \cite{Rajan} once
$|\theta_P(x)|$ becomes much larger than $T_{MO}$. Two distinct
types of phase diagrams were identified, one with $x_0 \geq x_K$
and the other with $x_0 < x_K$, which are schematically
represented in Fig.~\ref{G1}a. Both scenarios correspond to
different hybridization strengths ($\Gamma\propto T_K$): i) with
weak hybridization, i.e. of local moment (LM) character, and ii)
with moderate hybridization, i.e. of itinerant (ItM) character
\cite{S-FS}. This scheme applies to a doublet (N=2 degenerated)
ground state of Ce-$4f$ ions within the region where $T_{MO}(x)$
decreases. For $x > x_K$, $\Gamma \propto \theta_P$ increases more
rapidly with the consequent broadening of the magnetic levels.
Eventually, an overlap between the ground $\Gamma_{GS}$ and
crystal field (CF) excited levels may occur once respective
$\Gamma_{GS}$ and $\Gamma_{CF}$ hybridizations strengths become
comparable to the CF splitting ($\Delta_{CF}$). In the limit of
$\Gamma >> \Delta_{CF}$ the IV state, with N=6, takes over.

\begin{figure}
\begin{center}
\includegraphics[angle=0, width=0.5 \textwidth] {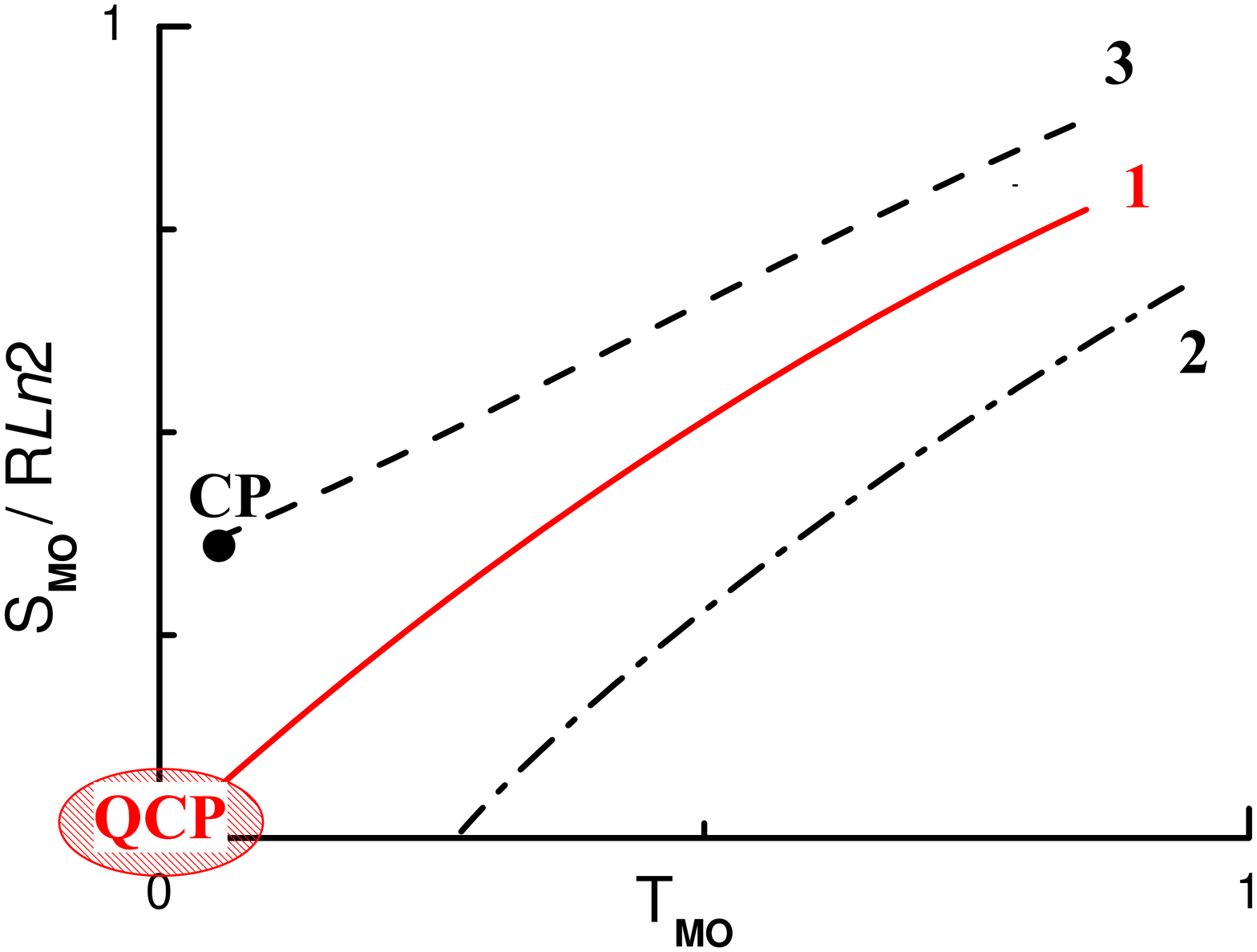}
\end{center}
\caption{Schematic representation of three possible dependencies
of $S_{MO}(T_{MO})$: 1) (solid line) fulfills the condition to
reach a QCP because $S_{MO}\to 0 $ as $T_{MO}\to 0$; 2) (dash-dot
line) a phase boundary vanishing at finite temperature and 3)
(dashed line) a phase boundary ending at a finite temperature
critical point (CP), see the text.} \label{G2}
\end{figure}

An encompassing phase diagram was proposed (see Fig.~\ref{G1}b)
computing the difference ($\Gamma_{GS} - \Gamma_{CF}$) versus the
sum ($\Gamma_{GS} + \Gamma_{CF}$) of those parameters. According
to experimental evidences from the temperature dependence of
electrical resistivity ($\rho(T,x)$), the LM regime correspond to
systems with $\Gamma_{GS} < \Gamma_{CF}$ and the ItM one to those
with $\Gamma_{GS} > \Gamma_{CF}$ \cite{sereni95}. Recently, some
non magnetic Ce compounds were reported to follow
Coqblin-Schrieffer model predictions for a four fold $N=4$ GS
\cite{Deppe,ScheidtGe4,CoqSchrieff}. Nevertheless, the limit
between a $N=4$ GS and a quasi-quartet (i.e. two doublets with
small but comparative $\Gamma_{GS} > \Gamma_{CF}$) are difficult
to be discriminated experimentally.

\section{Different Types of magnetic phase diagrams}

With the growing interest on quantum critical phenomena, a number
of theoretical models were proposed to describe low lying energy
excitations related to the $T_{MO}\to 0$ physics, see for example
Refs. \cite{HvL,Si,Coleman,Contin} and references therein. Since
the scope of the models is generally focused on microscopic
mechanisms and they are applied on specific exemplary cases, there
is an absence of an encompassing criteria able to detect or
discard new candidates and, eventually, to recognize new
alternative behaviors. Thermodynamic postulates provide the proper
tools for such a purpose because of their simplicity and
universality. Notably, the constraints imposed by the third law of
thermodynamics (e.g. the $T=0$ singularities in thermodynamic
parameters) are some times left aside.

Another relevant aspect concerns whether there is any condition to
be fulfilled at finite temperature for a real system to actually
reach the $T_{MO} \to 0$ limit. It is evident, for example, that
many $T_{MO}(p)$ phase boundaries are not properly checked to
involve the corresponding entropy of the ordered phase ($S_{MO}$)
despite their $T_{MO}(P)$ are na\"ively extrapolated from $T>1$\,K
down to e.g. their respective superconductive domes at $T_{sc} <
1$. Similar questions apply to concentration driven systems since
the arbitrarily monotonous extrapolations of $T_{MO}(x)$ exclude
any variation of $
\partial T_{MO}/\partial x$, even in the range where thermal and quantum
fluctuations start to compete in energy at low temperature.

The amount of intermetallic compounds claimed at present to be
candidates for quantum critical behavior allows, and even
requires, such a comparative analysis of their thermodynamic
properties in order to distinguish between reliable candidates and
potential ones. Furthermore, novel alternative behaviors to the
quantum phase transitions (QPT) might be missed due to the
mentioned simplistic extrapolations done in low temperature phase
diagrams.

\begin{figure}
\begin{center}
\includegraphics[angle=0, width=0.5 \textwidth] {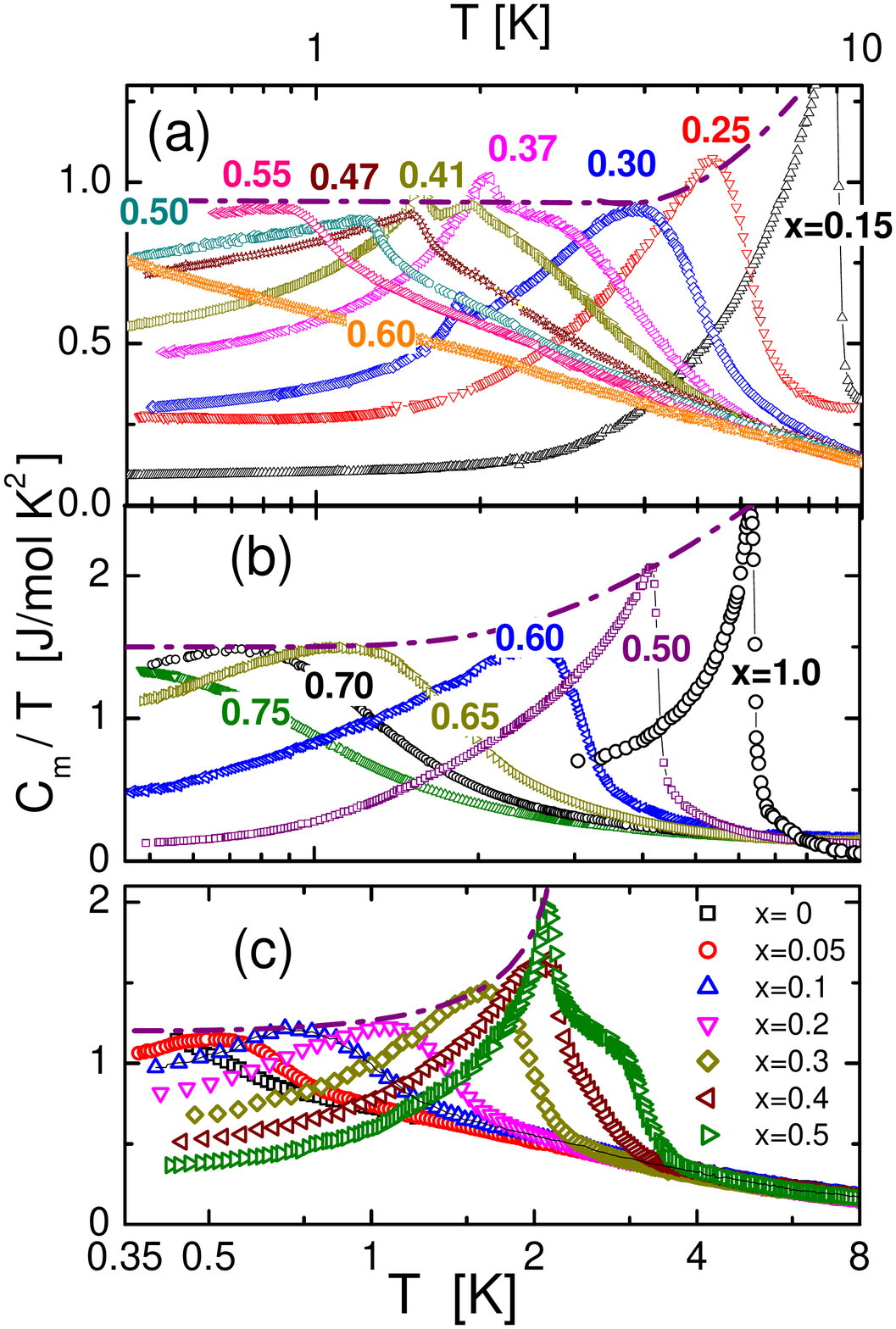}
\end{center}
\caption{(Color online) Magnetic contribution to specific heat
(after respective phonon subtraction) divided temperature of three
exemplary compounds approaching their respective critical regime.
(a) AF-CeIn$_{3-x}$Sn$_x$, (b) FM-CePd$_{1-x}$Rh$_x$ and (c)
AF-CePd$_2$(Ge$_{1-x}$Si$_x$)$_2$ after \cite{anivHvL}. Dash-dot
curves indicate the tendency to a nearly constant value of $C_m/T$
maxima within the $x_0 < x < x_{cr}$ region. Notice the
logarithmic $T$ dependence.} \label{G3}
\end{figure}

The simplest starting criterion for such an analysis is to take
into account all possible alternative scenarios for the decreasing
entropy of the magnetically ordered phase ($S_{MO}$, computed from
zero up to $T_{M0}$). There are three possible alternatives: 1)
$S_{MO}\to 0$ as $T_{M0}\to 0$; 2) $S_{MO}\to 0$ at finite
temperature due to an eventual transference of degrees of freedom
to a non-magnetic component, and 3) $S_{MO}$ does not decreases
proportionally to $T_{MO}$, producing a entropy accumulation at
low temperature. These three cases are schematically represented
in Fig.~\ref{G2} where each case is identified with the respective
number.

The exemplary systems selected for this study are Ce-lattice
compounds, where composition or chemical pressure effects are
produced by alloying Ce-ligand atoms. This criterion preserves the
periodicity and local symmetry of the Ce net in order to minimize
side effects such as disorder \cite{LMvsItM}. Hence, hereafter we
will refer to 'alloyed compounds' as those where only Ce-ligands
are doped or substituted. Otherwise, the few experimental results
extracted from a Ce diluted system will be explicitly indicated.

\subsubsection{Ce systems with $S_{MO}\to 0$ as $T_M\to 0$}

This is the thermodynamic condition to be fulfilled for reaching a
zero temperature QCP. For this group, we have selected some
exemplary Ce-lattice alloyed compounds whose specific heat were
measured down to very low temperature, see Fig.~\ref{G3}. The
common feature of these experimental results is that the
respective specific heat jumps ($\Delta C_m/T$) at $T_{MO}$ first
decrease and then broaden levelling off around $x\approx x_0$
\cite{anivHvL}. This change of regime at $x\approx x_0$ is
associated to a change of slope of the phase boundary at $x\approx
x_0$ whose implications are discussed in the following Section.
The $C_{m}/T$ maximum for $x \geq x_0$ can be analyzed within the
Ginzburg-Landau theory for second order transitions:
$C_m/T=a^2/2b$, where $a$ and $b$ are the coefficients of the free
energy expansion $G(\psi,T)=G_0(T)+ a(T)\psi^2+ b(T)\psi^4$. The
tendency to a constat value of $C_{max}/T$ indicates that the
$G(\psi,T)$ dependence on the $a^2/b$ ratio is locked and
consequently the entropy evaluated up to that maximum decreases
monotonously, following a sort of law of corresponding states
\cite{Callen} with the critical point at $T=0$. This is a {\it
necessary} thermodynamical condition for any system to reach a
QCP. Based on this analysis, one may include into this group the
well known system CeCu$_{6-x}$Au$_x$ \cite{Lohenysen} and
Ce(Pd$_{1-x}$Ni$_x$)$_2$Ge$_2$ \cite{KnebBrando} because they
exhibit the same fatures. one may even propose a potential
candidate like Ce(Pd$_{1-x}$Rh$_x$)In \cite{Bruck} that, to our
knowledge, was not yet investigated down to sufficiently low
temperature. Among the pressure driven systems, one finds CePb$_3$
\cite{CePb3press} showing this type of behavior up to 7GPa where a
change of magnetic structure occurs. Since this class of magnetic
phase diagrams are directly related to quantum critical phenomena,
we shall discuss in detail some selected experimental results in
Section III.

\begin{figure}
\begin{center}
\includegraphics[angle=0, width=0.5 \textwidth] {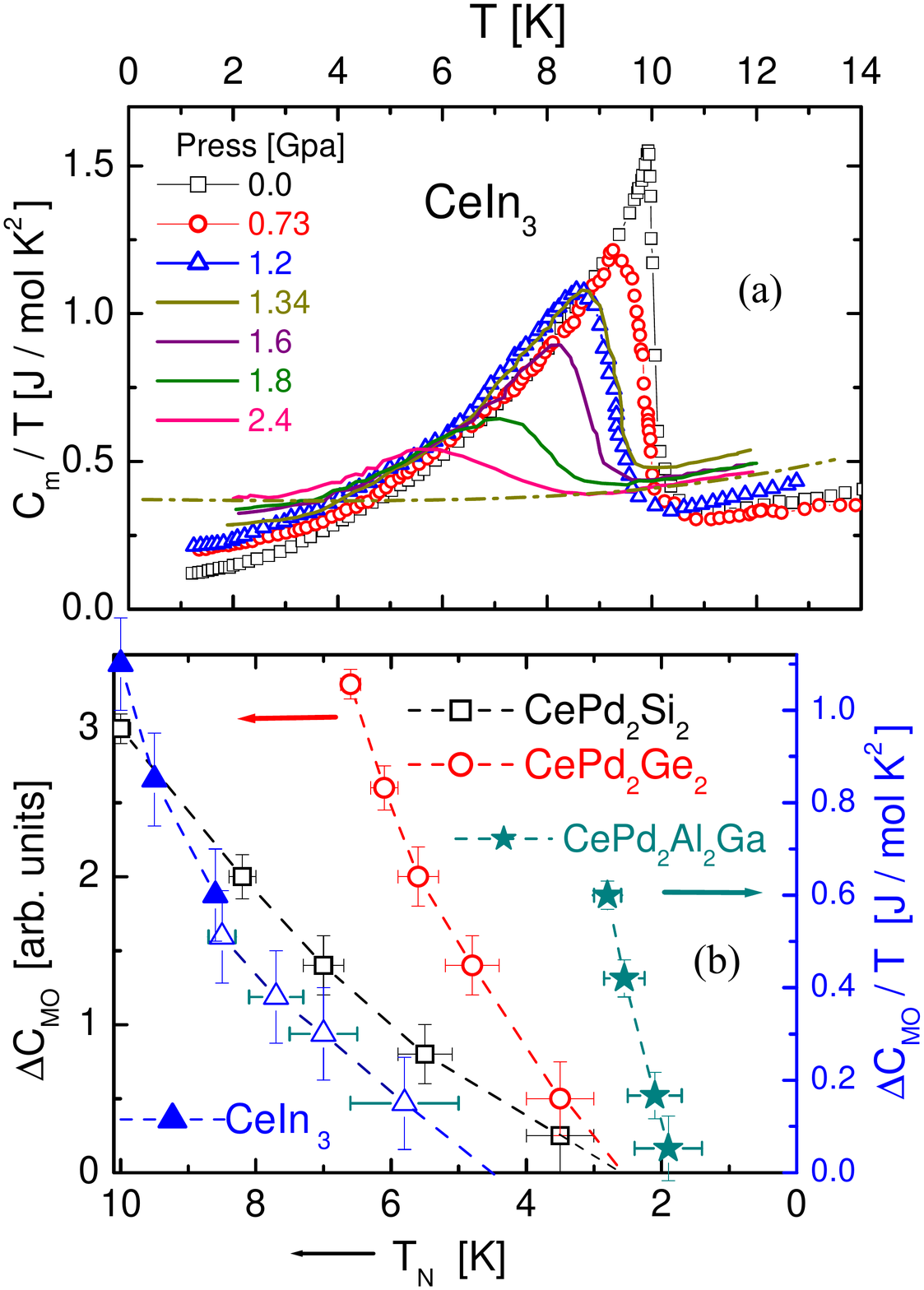}
\end{center}
\caption{(Color online) (a) Example of vanishing specific heat
jump at finite temperature as a function of pressure after
Ref.\cite{Peyrard} (open symbols) and Ref.\cite{Knebel}
(continuous lines). Dashed curve: non-magnetic HF reference for
e.g. 2.4\,GPa isobar. (b) Various vanishing specific heat jumps as
a function of $T_N(p)$. Notice that $T_N(p)$ decreases to the
right a pressure increases. Open symbols (left axis) extracted
from ac-specific heat and full symbols (right axis) from standard
heat pulse techniques, with CeIn$_3$ comparing both
techniques.}\label{G4}
\end{figure}

\subsubsection{Systems with $S_{MO}\to 0$ at finite temperature}

A second class of magnetic Ce-lattice systems behave quite
differently because $S_{MO}$ decreases faster than $T_N$
extrapolating to $S_{MO}=0$ at finite temperature, being all
experimental examples AF. This type of behavior is frequently
observed in pressure driven stoichiometric compounds and the
relevance of this class of phase diagrams arises from the fact
that many of them exhibit a superconductive GS under pressure
\cite{superc}.

The appearance of a superconductive dome is currently related to
the extrapolation of the AF phase boundary. However, a
thermodynamic analysis of those phase boundaries (mostly
constructed from resistivity measurements) reveal that such a
putative extrapolation of $T_N(p)$ down to $T_{sc}$ is quite
arbitrary. It is known that to extract absolute values from heat
capacity measurements under pressure exceeding 1.5\,GPa is quite
difficult. Nevertheless, measurements up to 1.2\,GPa, e.g. in
CeIn$_3$ \cite{Peyrard} and CePd$_2$Al$_2$Ga \cite{Eichler}, are
good reference for those performed at higher pressure, e.g.
CeIn$_3$ \cite{Knebel}, CePd$_2$Si$_2$ \cite{Umehara} and
CePd$_2$Ge$_2$ \cite{Bouquet}. These results
provide convincing information to recognize their distinct
behavior respect to those described in the previous subsection
(see Fig.~\ref{G3}). Since very high pressure results are given
using arbitrary units, a quantitative evaluation of $S_{MO}(p)$
variation is not possible. Alternatively one may evaluate the
relative decrease of the specific heat jump at $T_{MO}$ ($\Delta
C_{MO}(p)$)taking into account that $\Delta C_{MO}\to 0$ implies
that $S_{MO}\to 0$.

The common feature of these results is that they show a
progressive transference of the magnetic degrees of freedom to a
non-magnetic HF component in the region where $T_N(p)$ decreases.
As an example, the $C_m(T)$ results obtained on CeIn$_3$
\cite{Peyrard,Knebel} is presented in Fig.~\ref{G4}a. The
comparison with the other compounds is done using the relative
variation of $\Delta C_{M0}$ driven by pressure as depicted in
Fig.~\ref{G4}b. In that figure (right axis) the quantitative
comparison result is made using the results from CePd$_2$Al$_2$Ga
\cite{Eichler}.

The main conclusion extracted from Fig.~\ref{G4}b is that in all
these compounds the $\Delta C_m(T_N)$ jump vanishes at finite
temperature independently of their eventual superconductive GS,
with the phase boundary vanishing some degrees of temperature
above $T_{sc}$. In the case of CeIn$_3$, this behavior is
confirmed by $^{115}$In-NQR measurements \cite{NQR} under
pressure. Detailed electrical resistivity ($\rho$) measurements
performed on CePd$_2$Si$_2$ \cite{Demuer} also supports this
observation since the temperature derivative $\partial \rho/
\partial T$ (expected to be qualitatively related to the specific heat
\cite{Demuer}) shows a jump at $T=T_N$ which also vanishes at
finite temperature.

Noteworthy is the fact that these pressure driven superconductors
show a similar $T_N(0) \times p_0 = 2.9 \pm 0.2$\,KGPa product
\cite{Espanha}, being $T_N(0)$ the AF transition temperature at
ambient pressure and $p_0$ the pressure where the transition
vanishes. The origin of this empirical relation is not yet
elucidated, but it is confirmed by a number of non-superconductive
systems which do not verify that product. Among then, the
compounds undergoing a maximum of their phase boundaries as a
function of pressure (e.g. CePd$_2$Al$_2$Ga \cite{Eichler} or CePt
\cite{Larrea}) clearly do not present a superconductive phase.

It should be mentioned that the competition between magnetism and
superconductivity observed in CeRhIn$_5$ for example
\cite{Flouquet} cannot be included within this group because its
magnetic transition is of first order and there is a coexistence
of both phases below $T\approx 2$\,K and it merits its own
analysis.

\begin{figure}
\begin{center}
\includegraphics[angle=0, width=0.5 \textwidth] {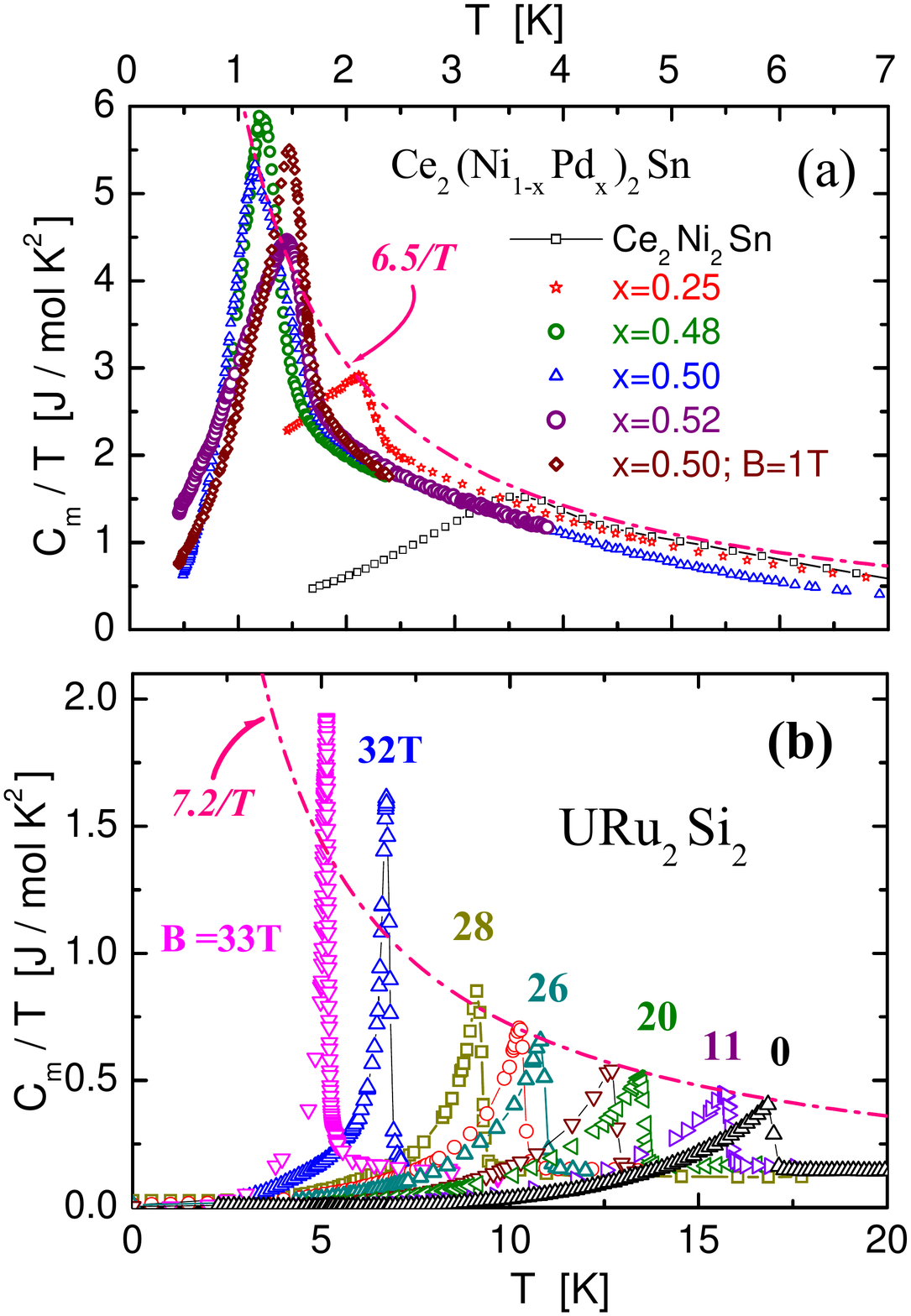}
\end{center}
\caption{(Color online) (a) Specific heat divided temperature of
the concentration dependent Ce$_2($Ni$_{1-x}$Pd$_x$)$_2$Sn system
after Ref.\cite{arXiv} and (b) referent field dependent
URu$_2$Si$_2$ after Ref.\cite{URu2Si2}} \label{G5}
\end{figure}

Concerning Ce-lattice alloyed systems, one can mention
Ce(Rh$_{1-x}$Ru$_{x}$)$_2$Si$_2$ \cite{CeRhRu2Si2},
Ce(Rh$_{1-x}$Pd$_{x}$)$_2$Si$_2$ and
Ce(Rh$_{1-x}$Ru$_{x}$)$_3$B$_2$ \cite{JPSJ2001} which show
equivalent vanishing process of their magnetically ordered degrees
of freedom. Interestingly, two ways of transfer can be
distinguished between ItM (large Fermi surface) and LM (small
fermi surface) magnetic systems. Within the former group (e.g.
Ce(Rh$_{1-x}$Pd$_{x}$)$_2$Si$_2$) the ordered state seems to build
up as a condensation of degrees of freedom from the $4f$ narrow
band heavy quasi-particles because an entropy compensation is
observed respect to a high temperature ($T > T_N$)
non-Fermi-liquid (NFL) component which does not change with
concentration \cite{JPSJ2001}. On the contrary, in the second type
the transference of degrees of freedom occur between two different
components because the NFL component increases at the expense of
the exhausting magnetically ordered ones. Preliminary $C_m(T)$
results on Ce(Co$_{1-x}$Fe$_x$)Si alloys clearly show this
transference between two systems because $\Delta C_m(T_N)$
vanishes around $T\approx 4.5$\,K with a coincident rising up of a
$C_m/T\propto -Ln(T/T_0)$ contribution. No superconductive GS is
expected in these alloyed systems because of the Ce neighbors
random distribution. Nevertheless, the non-magnetic stoichiometric
limit ($x=1$) of Ce(Rh$_{1-x}$Ru$_{x}$)$_3$B$_2$ shows low
temperature superconductivity \cite{Maple}.

\subsubsection{Systems with critical entropy accumulation as $T_N$
decreases}

\begin{figure}
\begin{center}
\includegraphics[angle=0, width=0.55 \textwidth] {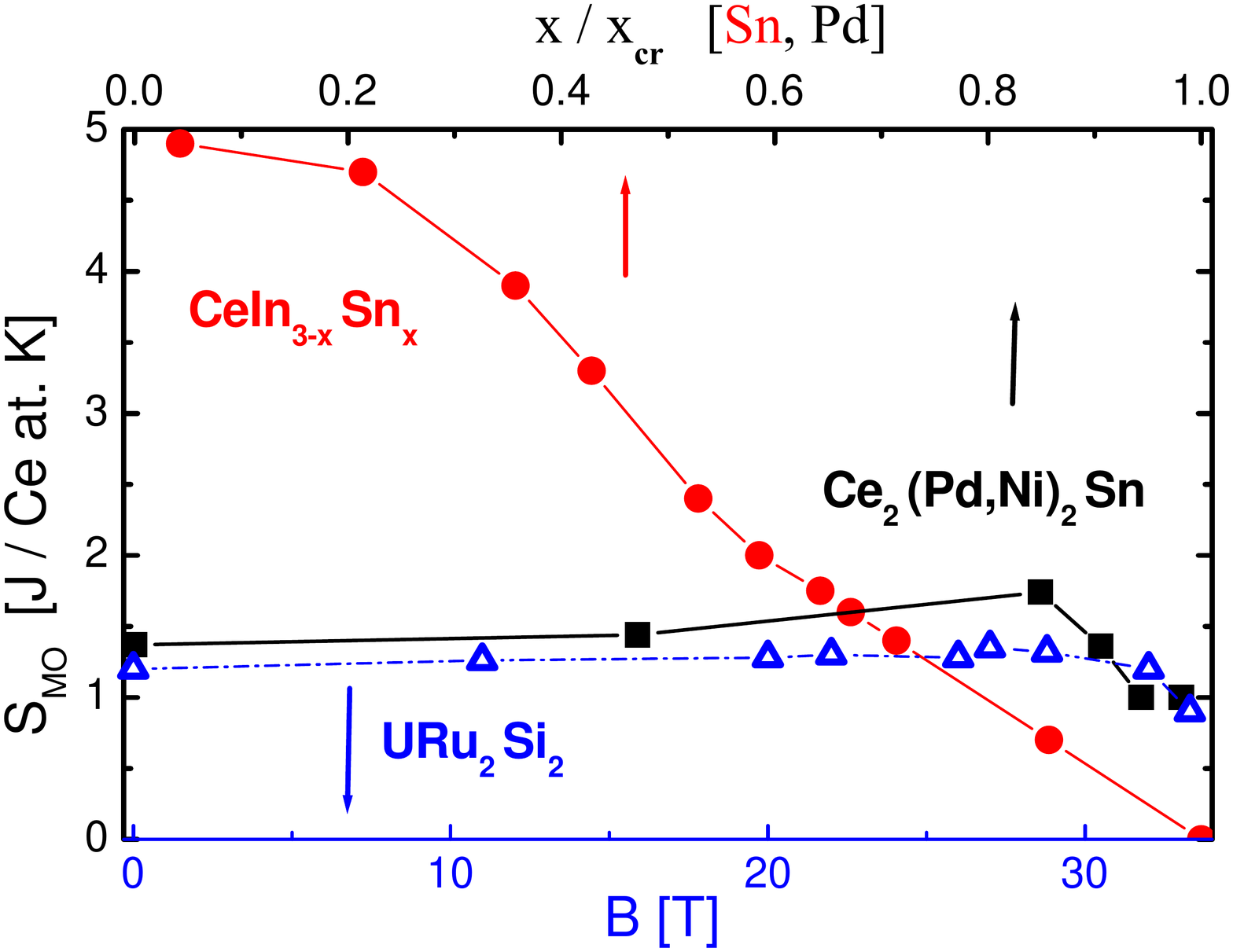}
\end{center}
\caption{(Color online) Comparison of the entropy gain $S_{MO}$ up
to $T_N$ between two types of behaviors:  CeIn$_{3-x}$Sn$_x$ of
the 1st. class and Ce$_2$(Ni$_{1-x}$Pd$_x$)$_2$Sn, and
URu$_2$Si$_2$ of the 3rd. class. Upper x-axis for concentration
$x$ dependent systems, and the lower x-axis for the magnetic field
dependent one. Notice that Ce$_2($Ni$_{1-x}$Pd$_x$)$_2$Sn contains
two Ce-at. per formula unit.} \label{G6}
\end{figure}

According to thermodynamics, if $S_{MO}$ does not decrease
proportionally to $T_{MO}(x)$ (like in the mentioned first class)
but with lower ratio, a $\lim_{T\to 0} S_{MO}
> 0$ would occur.
In that case, an entropy bottleneck occurs and the magnetic phase boundary shall end in a CP at finite
temperature where a first order transition drives the system to
$S_{MO}\to 0$. Such a situation is observed in the compounds
included in Fig.~\ref{G5}: Ce$_2$(Ni$_{1-x}$Pd$_x$)$_2$Sn
\cite{arXiv} and URu$_2$Si$_2$ \cite{URu2Si2}. The former is a
recently studied compound driven by Ce-ligands alloying, whereas
the latter is the well known U compound driven by magnetic field
and taken as a referent system for our purpose. It is worth to
note that the $C_m(x \,\rm{or}\, B)/T$ variation of the maxima are
described by the same function: 6.5 and $7.2/T$ respectively, both
drown in Fig.~\ref{G5}. In contrast to the behavior discussed in
subsection II-1 where $C_m/T_{max}$ becomes constant, in this case
is the $C_{max}(T_N)$ value that remains nearly constant till the
first order transition occurs (at $B\approx 33$\,T in
URu$_2$Si$_2$). The first order character of the transitions are
recognized from the value of the $C_m(T_N)$ maximum which clearly
exceed the $\propto 1/T$ function.

There are also striking coincidences in $S_{MO}$ concerning their
similar and nearly constant values that can be appreciated in
Fig.~\ref{G6}. The fact that different control parameters applied
on different systems produce the same effects can be taken as a
fingerprint for the universality of this behavior. For comparison,
in Fig.~\ref{G6} the $S_{MO}$ values obtained from
CeIn$_{3-x}$Sn$_x$, which belongs to the first group, are also
included to show that approaching the critical region the entropy
of these two compounds exceed that of one following the low of
corresponding states with a CP at $T=0$.

\section{Peculiar properties of Ce-lattice systems accessing to $S_{MO}\to
0$}

\begin{figure}
\begin{center}
\includegraphics[angle=0, width=0.5 \textwidth] {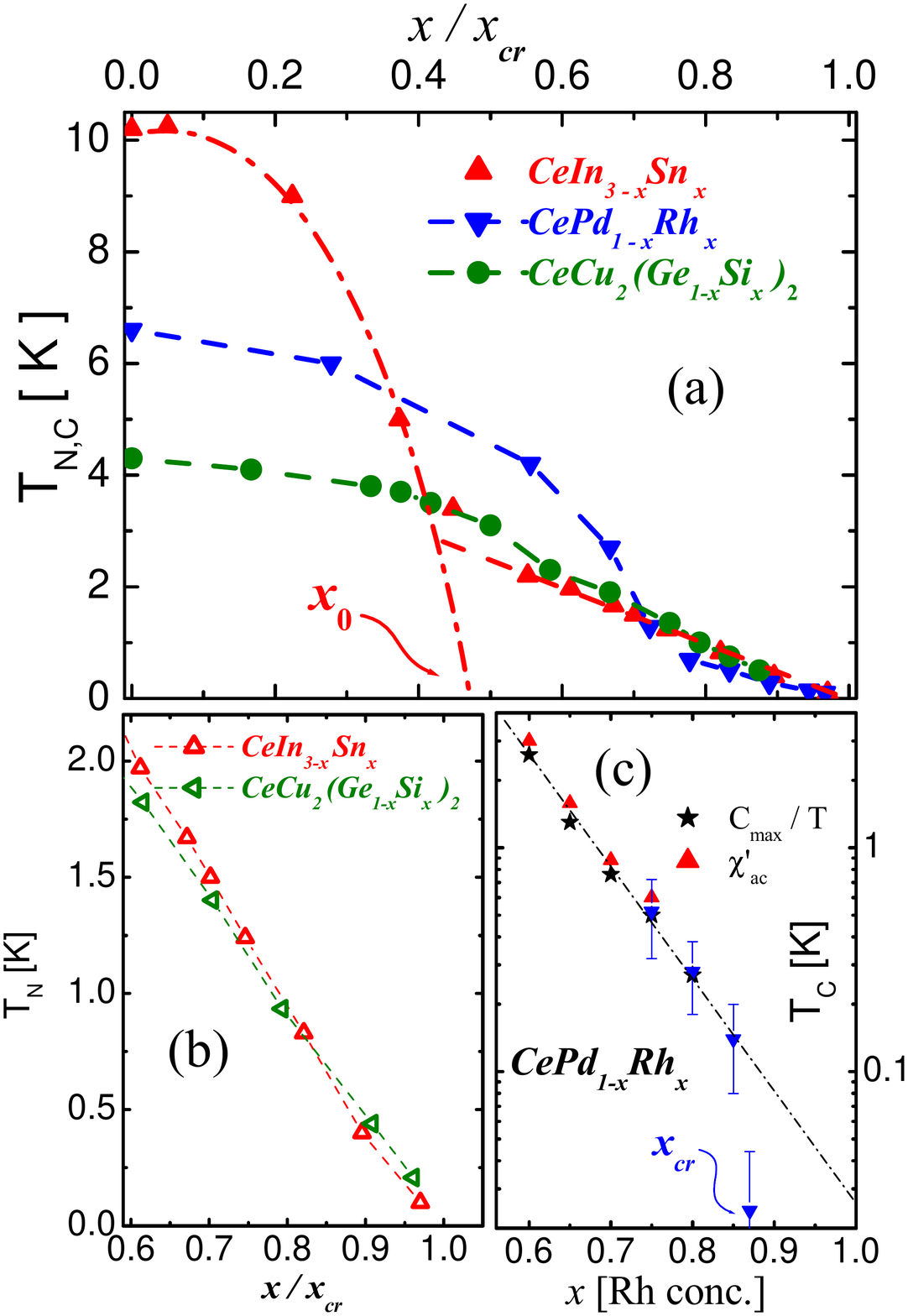}
\end{center}
\caption{(Color online) (a) Magnetic phase diagrams of three
exemplary compounds covering an extended range of temperature. (b)
and (c) show the detail of the quantum critical region for: (b)
two AF compounds with linear $T_N(x)$ dependence and (c) with an
asymptotic phase boundary of a FM compound \cite{CePdRh} (notice
the logarithmic $T_C$ axis).} \label{G7}
\end{figure}

At the time when the phase diagram presented in Fig.~\ref{G1} was
proposed, no quantum fluctuation effects were yet identified and
the usual low temperature limit for magnetic studies $\approx
1$\,K didn't provide evidences for such a scenario. Thus, the
$T_{MO}(x) \to 0$ limit was usually extrapolated following the
classical negative curvature to the concentration defined as $x_0$
in Fig.~\ref{G1}. Later on, lower temperature measurements made
evident that approaching $x_0$ a change of curvature (as presented
in Fig.~\ref{G7}) occurs around $T^{CR}\approx 2.5$\,K
\cite{FCM3}. Hence, the actual quantum critical concentration
$x_{cr}$ does not coincide with the $T_{MO}$ extrapolation to
$x_0$ but it occurs at higher values of $x$. Nevertheless, $x_0$
remains a relevant concentration because it characterizes the high
temperature region dominated by classical thermal fluctuations. To
our knowledge, this change of slope was not observed in phase
boundaries driven by pressure nor by magnetic field. However,
there is a striking coincidence in the fact that the phase
boundaries driven by pressure vanish at similar or higher
temperatures than $T^{CR}$.

Apart from the mentioned modification of the $T_{MO}(x)$
curvature, the change of regime at $x_0$ coincides with other
unpredicted features occurring around that concentration. Among
them, there is a first order transition $T_I(x)$ observed the
compounds included in Fig.~\ref{G3}, c.f. AF CeIn$_{3-x}$Sn$_x$
\cite{Pedraza} and CePd$_2$Ge$_{2-x}$Si$_x$ \cite{Octavio} and FM
Ce(Pd$_{1-x}$Rh$_x$) single crystals \cite{Micha}. The exemplary
case of CeIn$_{3-x}$Sn$_x$ is depicted in Fig.~\ref{G8} and will
be discussed in detail in Section III-B. Other Ce systems showing
a satellite first order transition in similar context are
Ce(Pd$_{1-x}$Rh$_x$)In \cite{Bruck} and
Ce(Cu$_{1-x}$Ni$_x$)$_2$Ge$_2$ \cite{Sparn}. All these transition
shows a similar sharpness, only depending of the quality of the
sample.

\subsection{Magnetic Phase Boundary within the Quantum Critical regime}

The observed change of regime can be explained taking into account
the competition between the decreasing energy of thermal
fluctuations and the temperature independent energy of quantum
fluctuations. While the classical transition extrapolates to
$x=x_0$ with coherent thermal fluctuations of the order parameter
decreasing with temperature, the critical fluctuation associated
to a QPT respond to a quantum-statistical description \cite{HvL}.
As mentioned before, only composition driven phase diagrams show
this clear change of regime at $T^{CR}K$ (see Fig.~\ref{G7}a). In
composition driven systems, the formation of static disorder
fluctuations (or "rare regions" \cite{TVojta}) was proposed to
explain the shift of the critical regime from $x_0$ to $x_{cr}$.

\begin{figure}
\begin{center}
\includegraphics[angle=0, width=0.6 \textwidth] {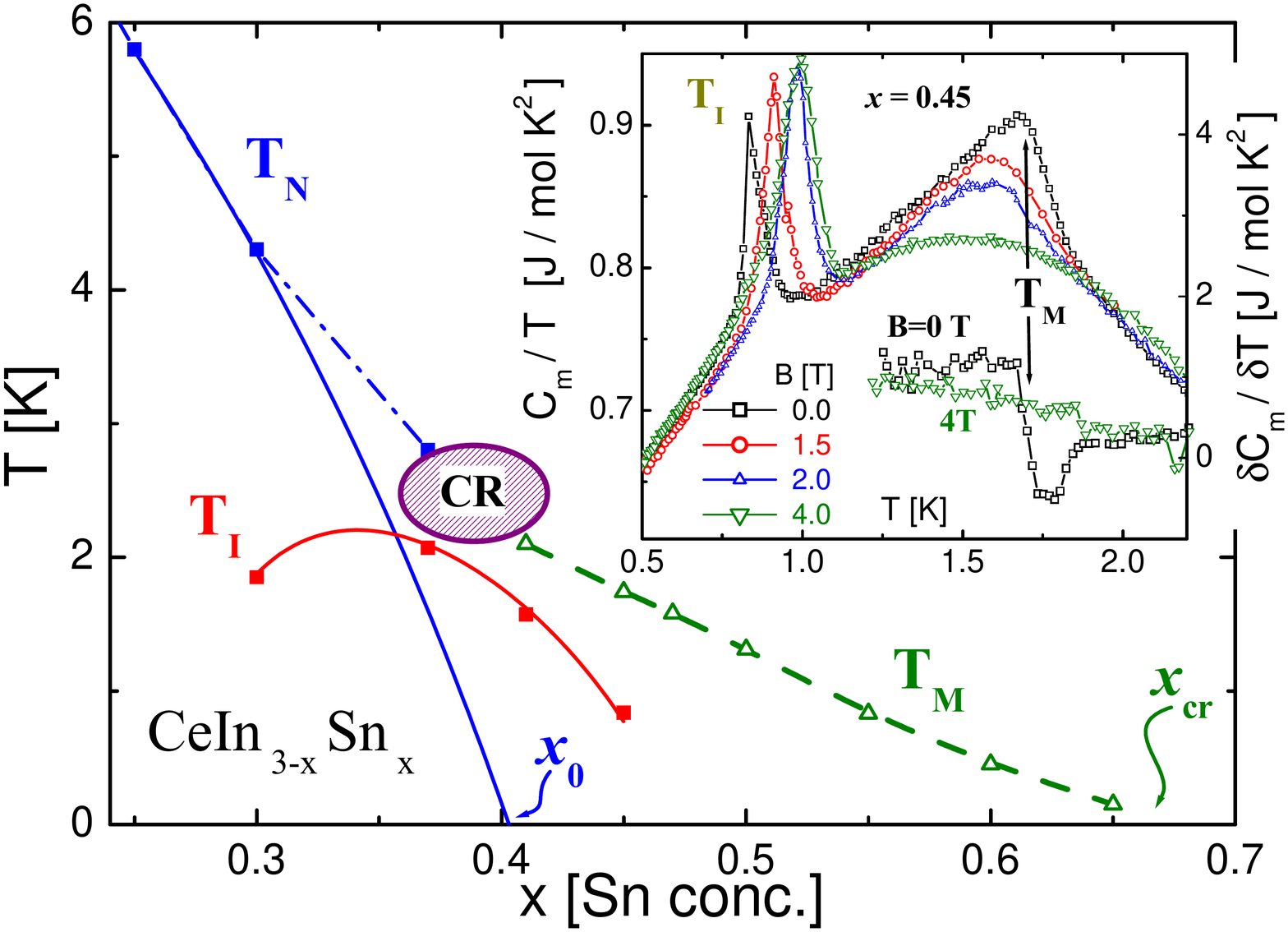}
\end{center}
\caption{(Color online) Detailed phase diagram of
CeIn$_{3-x}$Sn$_x$ around the critical region CR. One can identify
different phase boundaries: second order transitions $T_N$ for
$x\leq 0.3$, a first order dome $T_I$ around CR ($0.3\leq x \leq
0.45$) and the $T_M$ boundary within the pre-critical region
($x_0\leq x \leq x_{cr}$). Inset: Magnetic field effect on $T_M$
and $T_I$ transitions on sample CeIn$_{2.55}$Sn$_{0.45}$ after
Ref.\cite{Pedraza}. Right axis (lower curves) : suppression of the
$C_m(T_M)$ derivative jump for $B=4$\,T.} \label{G8}
\end{figure}

The occurrence of "rare regions" together with the so-called
Griffiths effects \cite{HvL} are mostly addressed to describe
composition driven systems because of their intrinsic possibility
of local disorder attributed to all alloyed systems. As it was
discussed in Subsection I-C, there is a clear difference between
{\it structural} and {\it chemical potential} pressure effects
because the former implies a random (i.e. disordered) Ce-ligand
atomic volume distribution. This effect cannot be na\"ivëly
extrapolated to chemical potential variation introduced by
neighbor elements with nearly equal atomic volume substitution.
This effect is therefore better described as an electronic
topological distribution. In fact, the change of regime occurs: i)
quite suddenly and only close to $x_0$, ii) in different compounds
with quite different relative concentrations and iii) without a
further broadening at higher concentrations \cite{Nordhiem}.
Moreover, the first order transitions at $T_I$ (see e.g. inset in
Fig.~\ref{G8}), show similar sharpness in all cases despite of
their different concentration regions excluding any atomic
disorder as a relevant factor.

All these experimental evidences confirm that there is a crossover
between two distinct regimes at similar temperature. Beyond that
region, QPT mechanisms dominate the low energy scenario producing
drastic changes in the nature of the magnetic phase boundary as it
will be analyzed in detail in the following subsection. Some of
those effects can be observed in Fig.~\ref{G7}b for two AF
compounds where $T_M \propto \mid x-x_{cr}\mid$ and in
Fig.~\ref{G7}c for a FM one, where $T_{C}(x)$ decreases
asymptotically till it collapses to zero at very low temperature.
Notice that for the $x>x_0$ region the phase boundary is label as
$T_M$ instead of $T_{MO}$ in order to distinguish them as
belonging to different regimes. If one describes these magnetic
phase diagrams within the pattern proposed in Fig.~\ref{G1}, one
may recognize that the systems included in Fig.~\ref{G7}b belong
to the $x_0<x_K$ class whereas that from Fig.~\ref{G7}c to the
$x_0 \geq x_K$ ones.

To gain insight into this unexplored range of concentration
between $x_0 \leq x \leq x_{cr}$, we will address our attention to
the phase diagram and the related thermal properties of
CeIn$_{3-x}$Sn$_x$. In Fig.~\ref{G8} we shown a detail of the
phase diagram around the critical region where three phase
boundaries converge: i) the classical AF-$T_N(x)$ dominated by
thermal fluctuations, ii) $T_M(x)$ dominated by quantum
excitations, and iii) the dome of a first order transition
$T_I(x)$ curve. The different nature of $T_M$ respect to the
classical AF-transition at $T_N(x<x_0$ can be clearly appreciated
in the inset of Fig.~\ref{G8} from the flattening of the
transition by magnetic field (up to $B=4$\,T) measured on sample
$x=0.45$. The same behavior is observed at higher concentration
down to the milikelvin temperature range \cite{Radu}. As a
comparison, one can mention that magnetic field applied to an
$x=0.25 < x_0$ sample \cite{colo} shows that a very high field
(about 40\,T) has to be applied to flatten the $T_N$ transition,
which then decreases in temperature before to vanish. Coming back
to the studied $x=0.45$ sample, the jump at $T_M$ can only be
observed analyzing the derivative of $C_m(T)$ as shown in the
lower part of that inset. To our knowledge, the CR scenario
presented in Fig.~\ref{G8} was only recently pointed out by theory
for itinerant FM systems \cite{Una}. However, our observations are
also include quite localized AF systems. Although a first order
dome occurs close to CR similarly as proposed by theory, the
following $T_M(x)$ transitions resamble Pippard's third order ones
\cite{Pippard}.

\subsection{Thermodynamic behavior of CeIn$_{3-x}$Sn$_x$ within
the $x_0<x<x_{cr}$ range}

\begin{figure}
\begin{center}
\includegraphics[angle=0, width=0.65 \textwidth] {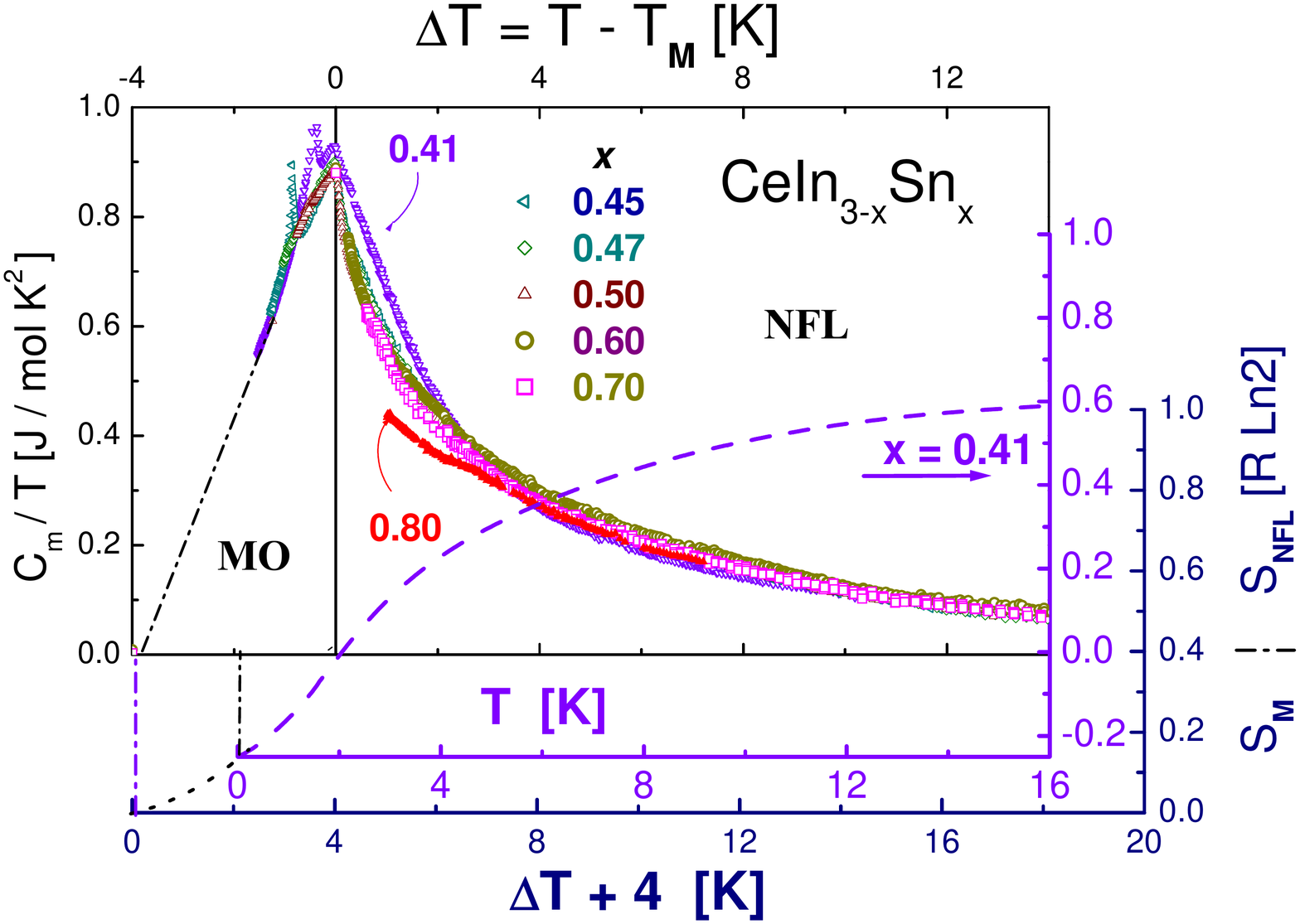}
\end{center}
\caption{(Color online) Overlap of $C_m/T (0.45\leq x \leq 0.70)$
curves plotted versus a shifted temperature $\Delta T=T-T_M$
(upper and left axes) after Ref.\cite{anivHvL}. The dashed curve
represents the concentration independent entropy gain for
$0.45\leq x \leq 0.70$ samples, discriminated between the $T<T_M$
range $S_{M}$ and the paramagnetic one $S_{NFL}$ (low $T$ and
inner $S$ axes). Full $RLn2$ entropy is computed from $\Delta
T=-\,4$K (lowest '$\Delta T+4$' and outer '$S_{M}+S_{NFL}$' right
axes), see the text.} \label{G9}
\end{figure}

In this subsection we will analyze the $C_m(T)$ and entropy
($S_m(T)$) contributions of the CeIn$_{3-x}$Sn$_x$ system, which
was investigated in detail within the $x_0<x<x_{cr}$ range
\cite{Pedraza}. Taking profit of the linear variation of $T_M$
with $x$, one can compare the $C_m(T)/T$ dependence of different
samples by normalizing the temperature as $\Delta T(x) = T -
T_M(x)$. In Fig.~\ref{G9} we show that comparison performed on
seven samples covering the concentration range between $0.41\leq x
\leq 0.80$. Notice that with this definition of $\Delta T$, the
magnetically ordered phase corresponds to the negative range of
that parameter, see the upper x-axis of Fig.~\ref{G9}. There, one
can see how the $C_m(T>T_M)/T$ tails of the alloys belonging to
the $0.45\leq x \leq 0.70$ range converge into a unique curve. In
order to remark the validity of this scaling, we have also
included in the figure the results obtained from $x=0.41$ and
$0.80$ samples, placed beyond the limits of the pre-critical
region, which clearly deviate from the scaled ones.

In Fig.~\ref{G9}, the vertical line at $\Delta T = 0$ line splits
the $C_m/T$ contribution into two parts, the $T<T_M$ one
(hereafter label as $C_{M}/T$) and the tail at $\Delta T > 0$
hereafter identified as $C_{NFL}/T$ because of its NFL behavior
(c.f. $\propto Ln(T/T_0)$). Notably, also the $C_{M}/T$
contribution for the samples within this pre-critical region
overlap each other at $\Delta T < 0$. Samples $x=0.41$ and 0.45
show the weak peak related to a first order transition discussed
in the previous subsection. The relevant feature is that the
$C_{M}/T$ overlap allows an extrapolation of $C_{M}/T \to 0$ to
$\Delta T \approx -\, 4$\,K which is also independent of
concentration. We remind that a $\Delta T < 0$ value does not
correspond to a negative temperature but simply to a common
extrapolation to a zero value of the $C_m /T$ contribution.

\begin{figure}
\begin{center}
\includegraphics[angle=0, width=0.6 \textwidth] {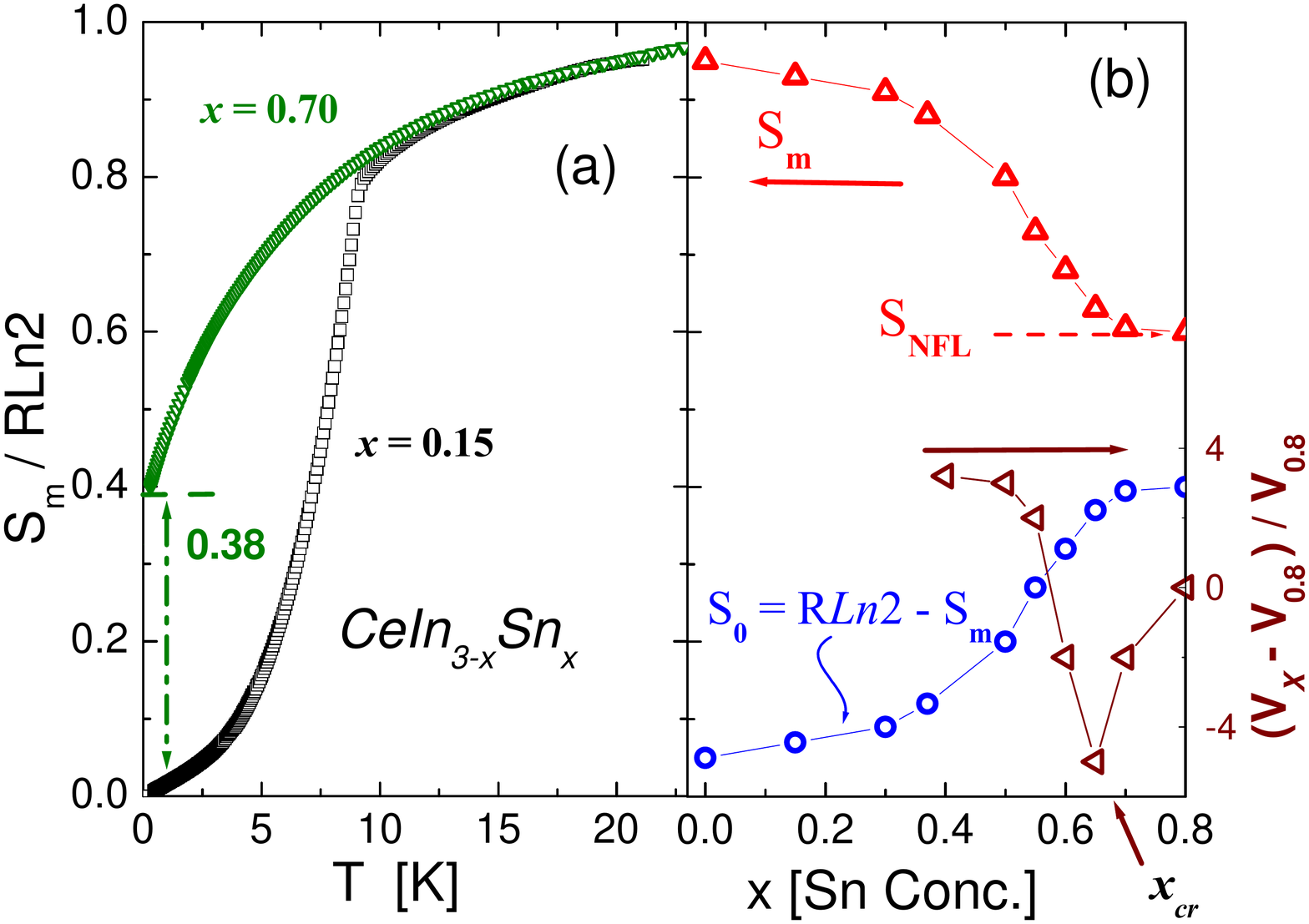}
\end{center}
\caption{(Color online) a) Comparison of the temperature
dependence of $S_m$ between the alloy $x=0.15$ belonging to the
classical region and one lying on top of the critical
concentration $x=0.70$ showing a deficit of $\approx 40\%$. b)
Left axis: concentration dependence of the entropy
$S_m=S_M+S_{NFL}$ of CeIn$_{3-x}$Sn$_x$ samples measured up to
20K, and the zero point entropy $S_0$ computed as the difference
respect the total expected value R$Ln2$. $x_{cr}$ highlights the
critical concentration. Right axis: relative change of volume at
$T\to 0$ for $x\geq 0.4$ samples, see text.} \label{G10}
\end{figure}

The key parameter to describe this peculiar behavior of the
specific heat is its associated entropy, evaluated as
$S_m=\mathbf{\int} C_m/TdT$. According to the definition proposed
for $\Delta T$, one may split the total measured values as
$S_m=S_{MO}+S_{NFL}$, being $S_{M}$ the contribution of the
$T<T_M$ phase (i.e. $\Delta T\leq 0$) and $S_{NFL}$ the one from
the NFL tail for $\Delta T \geq 0$. For the following analysis we
take as reference the entropy variation of sample $x = 0.41$
because it contains largest $S_{M}$ contribution among the samples
included in Fig.~\ref{G9}. As it can be appreciated in the figure,
the $S_{M}(x=0.41)$ contribution slightly exceeds $0.2RLn2$
whereas $S_{NFL}$ reaches $\approx 0.6RLn2$ (see inner right
axis). Noteworthy, the full $RLn2$ value is only reached once the
extrapolation to the $C_m/T = 0$ value at $\Delta T \approx -4$\,K
is included, as depicted using the lowest `$\Delta T+4$' and outer
`$S_{M}+S_{NFL}$' right axes in Fig.~\ref{G9}. Since $S_{NFL}
\approx 0.6RLn2$ does not change with concentration, but $S_{M}\to
0$ as $x \to x_{cr}$, one concludes that about $40\%$ of the
$RLn2$ entropy is missed as $T_M \to 0$. that difference is
illustrated in Fig.~\ref{G10}a by comparing the $S_m(T)$ variation
from samples $x=0.15$ (with full entropy) and $x=0.70 \approx
_{cr}$.

It is evident from Fig.~\ref{G9} that the decrease of $S_{M}(x\to
x_{cr})$ does not imply a transference of entropy to the NFL phase
because $S_{NFL}$ is independent of concentration within this
concentration range. Consequently, those degrees of freedom are
missed without any change of $S_{NFL}$, which remains unchanged
with a $60\%$ of the $RLn2$ value. We resume this situation in
Fig.~\ref{G10}b, where measured $S_m$ up to $T=20K$ and $S_0=RLn2
- S_m$ are represented as a function of concentration. At the QCP,
$S_m \to S_{NFL}$ because $S_M \to 0$ and $S_0 \to 40\% RLn2$.

Simplistic explanations looking for a some extra entropy
contribution at higher temperature fail because it would imply a
discontinuous transference of entropy from the $T<T_M$ to
temperatures above 20\,K according to Fig.~\ref{G9}. We recall
that in CeIn$_3$ the crystal-field excited quartet lies at high
enough energy ($\approx 100$\,K \cite{Lawrance}) which guaranties
no contribution to this analysis. Neither a Kondo temperature
increase can be argued because the temperature of the maximum of
the electrical resistivity ($T^{\rho}_{max}$) remains unchanged
between $x_0$ and $x_{cr}$ at $T^{\rho}_{max}\approx 19$\,K
\cite{Pedraza}.

The lack of entropy showed by this Ce compound is not an exception
because in the cases where this type of analysis was performed the
R$Ln2$ value for a doublet GS was never reached. For example, the
compounds showing a $C_m/T \propto Ln (T/T_0)$ dependence only
reach a $0.54 R Ln2$ value \cite{scaling}. Since the mentioned $Ln
(T/T_0)$ dependence corresponds to one of the possible scenarios
for QCPs predicted by theory \cite{HvL}, one infers that this
deficit in the entropy or the consequent arising of remnant
entropy at $T=0$ ($S_0$) is intrinsic to the NFL phenomenology
approaching that point.

The question arises whether the low temperature $x\to x_{cr}$
behavior is governed by low dimensional fluctuations as it was
observed in CeCu$_{5+x}$Au$_{1-x}$ \cite{HvL1999}. Unfortunately
neutron scattering studies on CeIn$_{3-x}$Sn$_x$ are not available
because of the strong neutron absorption of In nuclei.
Alternatively, one may check whether any evidence for low
dimensional fluctuations can be extracted from thermodynamical
results. For that purpose one can evaluate the internal energy,
$U_m(x)$ and $S_m(x)$ for $T_M < T < \infty$ in samples around
that concentration, and compare them with Ising and Heisenberg
model predictions \cite{20Domb} for 1, 2 and 3 dimensional systems
with different lattice structures (i.e. coordination number). Such
analysis showed that the those thermodynamic parameters nicely fit
into the predicted values for a 2D- Ising quadratic layer
\cite{FCM2}.

\subsection{Thermal Expansion}

In order to confirm that the anomalous evolution of the entropy
approaching the critical point as due to an intrinsic effect, a
complementary thermodynamic parameter sensitive to this phenomenon
has to be investigated. Such alternative is provided by thermal
expansion $\beta(T,x)$ measurements because they are related to
the entropy through the Maxwell relation $-\partial S/\partial P =
\partial V/\partial T = \beta$.Thus an anomalous $S_0(x\to x_{cr})$ dependence should have a
replica in $V_0(x\to x_{cr})$ as $T\to 0$. In this case, the
effective pressure is originated in the {\it chemical pressure}
produced by alloying.

The $\beta(T\to 0,x)$ variation of CeIn$_{3-x}$Sn$_x$ was studied
down to the mK range in the vicinity of the critical concentration
\cite{Kuechler}. In Fig.~\ref{G10}b we have included the volume
variation as $V_0(x)$ for $T\to 0$ extracted from the temperature
variation $V(T) = \int \beta dT$. The obtained values were
normalized well above any quantum fluctuation effect, i.e. $4\,K
\leq T \leq 8\,K$ taking as reference the $x=0.8$ alloy which lies
beyond the critical point. Both abnormal $S_0(x)$ and $V_0(x)$
dependencies are compared in Fig.~\ref{G10}b around $x_{cr}$.

\section{Thermodynamic implications of the $T\to 0$ physics}

\subsection{Third law of Thermodynamics and Remnant Entropy}

Thermodynamic postulates \cite{Callen} state that the entropy
decreases as $T \to 0$ reaching a finite value $S_0$, which is not
necessarily zero \cite{Pippard}. The implication of this postulate
can be viewed in a different way by considering the thermodynamic
definition of temperature \cite{Callen}, i.e. that $T=0$ is
reached once $\partial U/\partial S = 0$. Thus, the thermodynamic
condition for zero temperature corresponds to a zero variation of
the internal energy $U$ and not to the value of the entropy
itself. The $S_0=0$ value corresponds to a singlet GS without any
other accessible degree of freedom \cite{Abriata}. Meta-stable
states may eventually decay into such a GS in infinite time (like
e.g. amorphous or other structurally disordered systems), however
frustrated systems or those dominated by quantum fluctuations
(like those involved in the present study) escape to this
possibility.

The way to prove that $S_0 \neq 0$ was applied in Section III-C by
taking as reference the $S_m = RLn2$ value associated to the
magnetic doublet-GS of CeIn$_{3-x}$Sn$_x$ at $T\approx 20$\,K.
Interestingly, the absolute reference for the the entropy value is
taken form high enough temperature where both levels of the
doublet GS are equally occupied (i.e. $S_m=RLn2$).

A complementary aspect regarding the application of the third law
of thermodynamics implies the $\lim_{T\to 0} \partial S/\partial
T$. The $\partial S/\partial T =0$ possibility at $T=0$
corresponds to the previously mentioned case of a singlet GS,
realized in a long range ordered state and associated to a
positive $S_m(T)$ curvature (i.e. $\partial^2S_m /\partial
T^2>0$). The simplest examples for a $\partial S/\partial T \neq
0$ (or $\partial^2S_m /\partial T^2=0$) is provided by metallic
systems whose conduction electrons are described by standard
Fermi-liquid behavior with $\partial S/\partial T = \gamma$, c.f.
the Sommerfeld coefficient. Heavy fermion (HF) systems simply
increases the $\partial S/\partial T$ slope without changing their
physical properties.

Since a negative $S_m(T\to 0)$ curvature (i.e. $\partial^2S_m
/\partial T^2 < 0$) is not possible because it would imply a
singularity at $T=0$, the question arises whether there is an
upper limit for the $\partial S/\partial T = C_m/T$ slope or,
equivalently, for the $\gamma$ value in non ordered HF. Such a
question is related to the divergence of thermal parameters like
specific heat or thermal expansion because even a logarithmic
divergence at $T=0$ implies an infinite slope of $\partial
S/\partial T (T\to 0)$. This scenario was proposed by theoretical
models \cite{HvL} and claimed to correspond to experimental
results \cite{Stewart} dismissing some thermodynamic postulates.

\begin{figure}
\begin{center}
\includegraphics[angle=0, width=0.6 \textwidth] {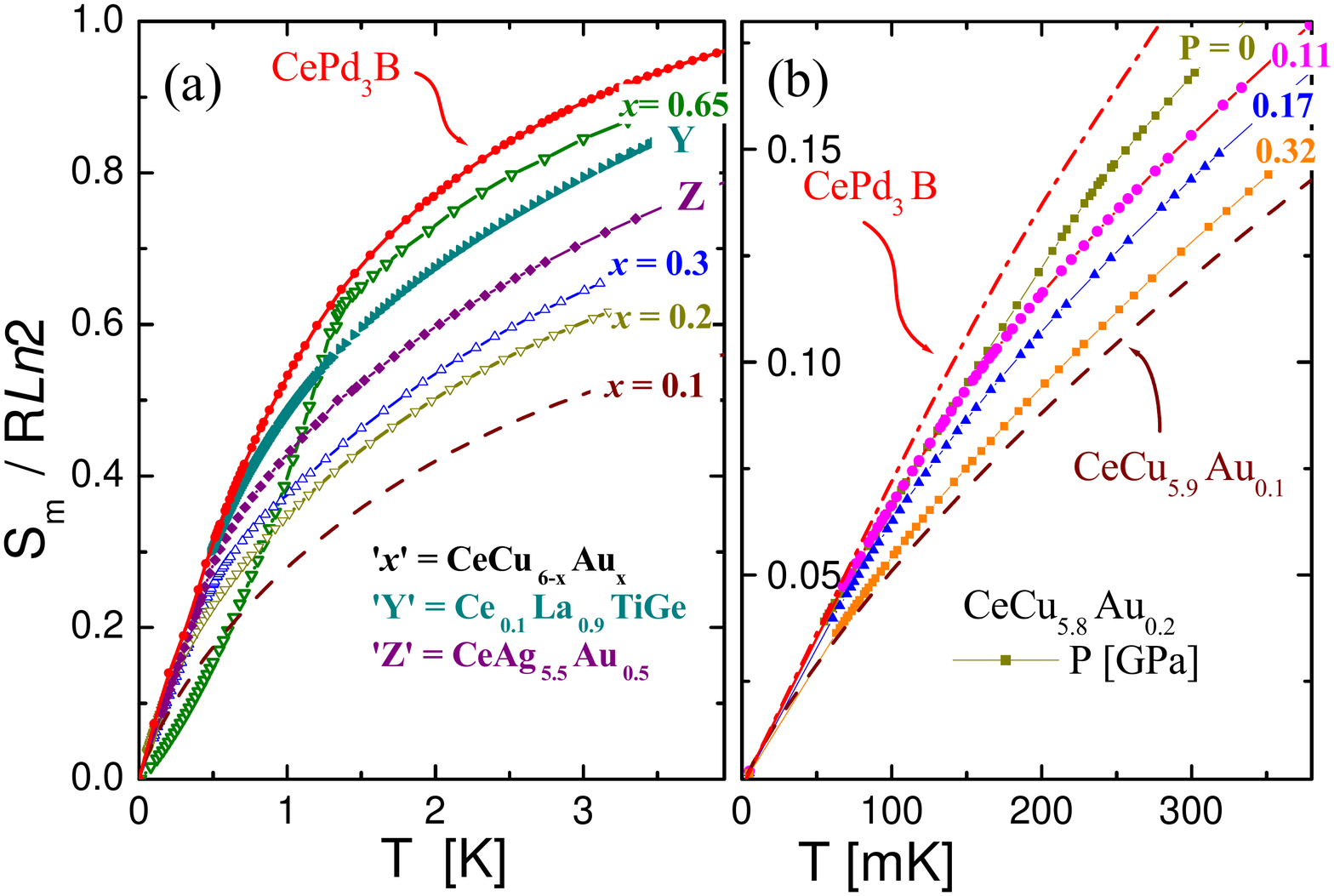}
\end{center}
\caption{Examples of maximum slope of $S_m(T)$ at $T\to 0$ in real
systems. (a) CePd$_3$B \cite{CePd3B}, CeCu$_{6-x}$Au$_x$
\cite{Lohenysen,13schl93}, (Ce$_{1-y}$La$_y$)TiGe \cite{CeLaTiGe}
and CeAg$_{6-z}$Au$_z$ \cite{ScheidtAg}. (b) Detail at the mK
range including: CeCu$_{5.9}$Au$_{0.1}$ (dashed line),
CeCu$_{5.8}$Au$_{0.2}$ at different pressures and, for comparison,
the fit for CePd$_3$B from $T > 0.5$\,K (dot dashed line).}
\label{G11}
\end{figure}

We will analyze now the existence of an eventual upper limit for
$\partial S/\partial T (T\to 0)$ and, in the following subsection,
the thermodynamic consequences of the proposed divergencies at
$T\to 0$. In Fig.~\ref{G11} we have collected the low temperature
$S_m(T)$ dependencies extracted from a number of Ce system showing
the highest values of $C_m/T\geq 3$\,J/molK$^2$ for a doublet GS
($N=2$) independently that they order or not. This comparison
includes the well known CeCu$_{6-x}$Au$_x$ \cite{Lohenysen} and
CeCu$_{6-x}$Ag$_x$ \cite{ScheidtAg} Ce-lattices. Also the diluted
(Ce$_{0.1}$La$_{0.9})$TiGe \cite{CeLaTiGe} is included to confirm
that this limit is related to thermodynamic properties
independently of a lattice configuration. Other Ce diluted
systems, like (Ce$_{1-x}$La$_x$)Pt$_3$Si \cite{CePt3Si}, also
exhibits similar high values for $0.8 \leq x \leq 0.95$.

As it can be seen, all experimental results show a sort of
envelope curve described by the $S_m(T)$ dependence of CePd$_3$B
\cite{CePd3B}. To our knowledge, the record of low temperature
entropy was extracted from the Ce-diluted
(Ce$_{0.03}$La$_{0.97})$B$_6$ \cite{satoB6} which exceeds the low
temperature $S_m(T)$ values of CePd$_3$B even after normalized by
$RLn4$ (remind that CeB$_6$ has a N=4 GS). Among the Ce-lattice
systems, CeNi$_9$Ge$_4$ shows a very high value: $C_m/T (T\to 0) =
5.2$\,J/molK$^2$ \cite{ScheidtGe4}, because its GS is composed by
the contribution of two hybridized doublets with a Kondo
temperature equivalent to the extremely small CF splitting
($\approx 10$\,K). Also in this case the corresponding
normalization entropy is $RLn4$.

From these experimental evidences we conclude that there is an
upper limit for $\partial S/\partial T (T\to 0) \approx
4$\,J/molK$^2$ for $N=2$ GS in Ce systems and consequently a lower
limit for the Kondo temperature evaluation based on the thermal
dependence of the entropy. The question arises whether this is a
physical limit for $T_K$ or simply a limit for the application of
theoretical models.

\subsection{Divergencies at $T \to 0$}

Two types of divergencies for thermodynamic parameters were
proposed to account for the low energy excitations in NFL systems
\cite{Stewart}, one described by a power law like $C_m/T \propto
T^q$ and the other by a logarithmic dependence like $C_m/T \propto
-ln(T/T_0)$. Details concerning the range of applicability of
different models exceed the scope of this phenomenological work
and can be found in many interesting review articles
\cite{HvL,Si,Coleman,Contin,HvL1999,Miranda}.

Most of Ce systems obeying a $C_m(T)/T$ power law dependence are
described by non fractional values of the exponent $q$, with
scarce coincidence with theoretical predictions devoted to systems
with short range order parameter fluctuations
\cite{Miranda,Contin}. Divergent power laws imply a non analytical
singularity at $T=0$ which is in conflict with the third law of
thermodynamics. In agreement with thermodynamics, actual specific
heat results show a systematic tendency to saturation of
$C_m(T)/T=\gamma_T$ at the low temperature limit, which is well
described by an heuristic modified power law $\gamma_T=G/(T^q+A)$
\cite{anivHvL}. In that formula $A$ represents an energy scale
below which the third law constraint of $C_m\to 0$ with $T\to 0$
\cite{Callen} becomes dominant. That parameter also allows to
evaluate the temperature of the crossover between NFL and FL
regimes since the later does not diverge at $T\to 0$. The $T=0$
limit of that formula is obtained computing the limit of
$\gamma_{T\to 0} =G/A$. In the case of the CePd$_3$B compound,
proposed in Fig.~\ref{G11} as the phenomenological envelope curve,
the fitting function obtained between 0.5 and 4\,K, is $C_m/T=
4.3/(T^{1.8}+1.1)$ which results in a $\gamma_{T\to 0} \approx
4$\,J/molK$^2$ value.

The other usual divergence observed in NFL systems has a
logarithmic character. Besides the fact that this type of
algebraic divergence does not imply a singularity at $T=0$, even
the system showing the highest $C_m/T$ values observed at low
temperature in CeCu$_{5.9}$Au$_{0.1}$ \cite{HvL1999} (included in
Fig.~\ref{G11}) does not exceed the $S_m(T)$ curve represented by
CePd$_3$B significantly. The comparative study performed
normalizing the temperature with the respective energy scale $T_0$
as $t=T/T_0$ \cite{scaling} shows that a universal function $C_m/t
= - D Log(t) + E_0 \times T_0$ describes all analyzed compounds,
with $D = 7.2$\,J/molK and $E_0$ accounting for any eventual high
temperature contribution (which is zero for
CeCu$_{5.9}$Au$_{0.1}$). Once subtracted the non logarithmic
contribution $E_0 \times T_0$ the computed entropy does not exceed
$60\%$ of the $RLn2$ value.

Interestingly, there is a common feature in systems showing a
$C_m/T \propto -Ln(T/T_0)$ divergency before to reach the critical
point, that is the specific heat jump transformed into a kink.
Such an anomaly is well illustrated by CeCu$_{5.8}$Au$_ {0.2}$ and
CeIn$_{2.55}$Sn$_{0.45}$ samples \cite{anivHvL}, which requires of
the temperature derivative of $C_m$ to observe a a discontinuity
at $T_m$ as shown in the inset of Fig.~\ref{G8}. That feature is
also observed in magnetic field driven phase boundary of
Sr$_3$Ru$_2$O$_7$ \cite{Rost}. Based on the fact that in
CeCu$_{6-x}$Au$_x$ the low energy excitations were recognized from
neutron scattering measurement to arise from magnetic fluctuations
with an effective dimensionality smaller than three
\cite{Schroeder}, one may infer that the mentioned cusp reflects
the low dimensionality of the order parameter in this quantum
critical region. Analyzing the $C_m(T)$ dependence around that
transition, its symmetry respect to $T_M$ reminds those observed
in other Ce compounds in similar conditions, i.e. with $T_M \leq
2.5$\,K \cite{FCM3}.

Thermal expansion also provides phenomenological information for
this scenario from measurements performed on
Ce(Pd$_{1-x}$Cu$_x$)$_2$Si$_2$ in two perpendicular crystalline
directions: $\alpha_c$ and $\alpha_{ab}$ \cite{Kromer}. In this
case one observes that, once the $\Delta C_m(T_N)$ jump
(characteristic for a 3-D mean field like transition) is smeared
by the magnetic decoupling between Ce planes, the corresponding
$\alpha(T)$ discontinuity transforms into a cusp before to vanish
as a function of $x$. The outstanding aspect of this change is
that it occurs at different concentrations depending on the
measured direction. While in 'c' direction the evolution goes hand
by hand with the specific heat jump vanishing at $x=0.2$, along
the 'ab' plane it holds up to $x=0.3$  but transforming into a
cusp at $T\approx 1$\,K as expected for a lower dimensionality
order parameter, see Fig.10 in Ref.\cite{JPSJ2001}.

\subsection{Consequences on the lower $T_K$ limit determination}

Further consequences arise from thermodynamical constraints
imposed by the third law concerning the evaluation of the Kondo
temperature from the entropy, e.g. $S_m(T_K) = 2/3 RLn2$
\cite{Desgr}. If there is an upper limit for the $\partial
S_m/\partial T$ derivative, a consequent lower limit occurs on the
value of $T_K$ extracted from $S_m(T)$. From the envelope curve
proposed in Fig.~\ref{G11} such a limit would be $\approx 1.3$\,K.
Similar situation occurs with models extracting $T_K$ from the
specific heat jump at $T_N$ as $T_K/T_N \propto \Delta C_0/\Delta
C_m$ \cite{Braghta}, being $\Delta C_0 = 1.5R$ the reference value
from mean field calculation of $\Delta C_m$ for a doublet GS. In
this procedure the fixed 1.5R value contradicts the constraint
imposed by the mentioned law of corresponding states which
requires that $\Delta C_m\to 0$ as $T_N\to 0$ affecting the
application of the $\Delta C_0/\Delta C_m$ ratio for a $T_K$
evaluation. Another current criterion to evaluate the Kondo
temperature is to compute $T_K \propto 1/\gamma_0$ \cite{Rajan}.
Also in this procedure the minimum value of $T_K$ is limited by
the empirical maximum of $\lim_{T\to 0}\partial S_m/\partial T =
\gamma_0$ observed in Ce systems. Whether this low $T_K$ limit
arising from thermodynamic conditions on $\gamma_0$ and $\Delta
C_m$ is intrinsic to the Kondo effect itself is an open question.
Theoretical models remark the possibility of a quenching of Kondo
effect approaching a QCP \cite{Si,Coleman}, but not based on
thermodynamic constraints. In any case, these considerations warn
on the accurate application of theoretical models not accounting
for thermodynamical constraints on real systems.

\section{Conclusions}

This comparative analysis of the low temperature properties of
Ce-magnetic systems show the power of thermodynamic parameters in
recognizing different types behaviors, in particular the not fully
profited information extracted from the entropy at $T\to 0$.
Moreover, the third law of thermodynamic provides universal {\it
sine quibus non} conditions for real systems to approach a zero
temperature QCP, independently of {\it a priori} model hypothesis.
It is important to distinguish between candidates to present QCP
and other with alternative behaviors because new relevant physical
phenomena might be missed due to a misleading low temperature
extrapolation.

A significant amount of experimental evidences were analyzed at
the light of these criteria, which allow to conclude that at least
three types of phase diagrams can be clearly distinguished.
Depending on the behavior of the $T_{MO}$ phase boundaries, those
phase diagrams can be sorted as follows: i) those where the phase
transition fulfills the conditions to be driven to $T=0$, ii)
those whose phase boundaries vanish at finite temperature because
their MO degrees of freedom are progressively transferred to a non
magnetic component, and iii) those ending in a critical point at
finite temperature.

In the first case the possibility to reach a QCP is supported by
the continuous decrease of the $S_{MO}$ entropy, which
extrapolates to zero as $T_{MO}\to 0$. Despite of its monotonous
decrease, the phase boundary driven by alloying Ce-ligands shows a
change of curvature at $x=x_0$. This behavior is attributed to a
change from classical to quantum critical of regime since beyond
that concentration quantum fluctuations seem to dominate the
scenario. Strikingly, such a change occurs at similar temperature
energy $T^{CR} \approx 2.2\pm 0.3$\,K in all studied systems, and
below that temperature a tendency to saturation of the maximum of
$C_m(T)/T$ arises as a distinctive characteristic according to a
law of corresponding states for a $T=0$ critical point.

A number of distinctive properties were highlighted by a detailed
analysis of the thermal properties of the exemplary system
CeIn$_{3-x}$Sn$_x$: i) there is an anomalous reduction of about
$40\%$ of the entropy respect to reference value $R\ln2$ expected
for a doublet GS. This missed entropy can be regarded as a zero
temperature remanent entropy, ii) it can be quantitatively
demonstrated that, contrary to current suppositions, the reduction
of $S_{MO}$ as $T_{MO}\to 0$ is not transferred to the
paramagnetic phase, and iii) at the critical region defined by
$x=x_0$ and $T=T^{CR}$ there is a systematic presence of a first
order transition and beyond that point the phase boundary $T_M(x)$
changes its nature, manifested in a strong dependence on magnetic
field and a $C_m(T)$ jump only observed in its temperature
derivative.

No evidences for $T\to 0$ divergencies are observed in real
systems, instead a progressive saturation of $C_m/T$ is observed
in those cases described by a power law a finite temperature.
Neither those systems with $\propto -Ln(T/_0)$ dependence exceed
the empirical upper limit of $\gamma_0 \approx 4$\,J/molK$^2$ for
Ce systems with doublet GS.

The second type of $S_{MO}(T_{MO})$ behavior is currently observed
in pressure driven phase boundaries. In this case, specific heat
results indicate that the phase boundary itself vanishes because
of a progressive transference of degrees of freedom to the
non-magnetic component occurring at $T\geq 2$\,K. Despite of the
formation of a superconductive phase their magnetic phase
boundaries do not reach that transition because it occurs below
the 2\,K threshold. This type of behavior is also observed in
Ce-ligand alloyed system, but there the occurrence of
superconductivity is unlikely because of allying effects.

The distinct characteristic of the third class of phase diagrams
is given by the fact that the $C_m(T_N)$ maxima values are found
to be constant instead of $C_m/T$ like in the first group. In this
case the entropy accumulation as $T_N$ decreases makes the phase
boundary to end at a finite temperature critical point. There, a
first order transition drops $S_{MO}$ down to 0. This peculiar
scenario was detected in a system driven by Ce-ligand composition
and confirmed by a well know U compound driven by magnetic field.
Notably both systems coincide in their $S_{MO}$ values.

To our knowledge, many of these experimental observations were not
predicted by current models focused into the physics of QCPs. This
is probably due to the difficulty of a generic treatment of
thermodynamic parameters like entropy or the specific heat jump in
a region dominated by a complex spectra of quantum excitations.

\section*{Acknowledgments}
The author acknowledges A. Eichler, M. Jaime, G. Knebel, M. Deppe
and R. Kuechler for allowing to access to experimental data, to S.
Grigera, U. Karahasanovic and C. Proetto for illustrative
discussions. The studies on CeIn$_{3-x}$Sn$_x$ and
CePd$_{1-x}$Rh$_x$ systems were carried in collaboration with C.
Geibel supported by DAAD, Alexander von Humboldt Fundation,
PICTP-2007-0812 and SeCyT-UNCuyo 06/C326 projects. Experimental
contribution of M. G\'omez Berisso and P. Pedrazzini is also
acknowledged.


\begin{thebibliography}{99}
\bibitem{S-FS} These types of GS are currently identified as related to small (LM) or large (ItM) Fermi-surfaces, see for example \cite{Si}.
\bibitem{LMvsItM} Strictly, long range of short range magnetic
order are not different in the following dicussion becauseit is
base on the entropy (i.e. degress of freedom) involved into the
decreasing magneic interactions.
\bibitem{Nordhiem} Remind that Nordheim criterion \cite{Mott} for disordered
alloys propose a $\rho_0 \propto x(1-x)$ distribution, which is
centered at $x=0.5$. An asymmetrical $x$ dependence is expected
for alloyed elements with different electronic orbitals, including
$s-d$ electronic transition.
\end{thebibliography}
\end{document}